\documentstyle[preprint,aps,epsfig]{revtex}
\draft
\begin{document}               
\title{Rotating ground states of trapped 
Bose atoms with arbitrary two-body interactions
}
\author{ O.K. Vorov$^1$\footnote{vorov@ganil.fr},  
M. S. Hussein$^2$ and P. Van Isacker$^1$
}
\address{
$^1$GANIL, BP 5027, F-14076,   Caen, Cedex 5, France \\
$^2$Instituto de Fisica, 
Universidade de Sao Paulo \\
Caixa Postal 66318,  05315-970,  \\
Sao Paulo, SP, Brasil
}
\date{12 July 2002}
\maketitle
\begin{abstract}
In a $k$-dimensional system of 
weakly interacting Bose atoms
trapped by a spherically symmetric and harmonic external potential,
an exact expression is
obtained for the rotating ground states
at a
fixed angular momentum.
The result is
valid for arbitrary interactions obeying minimal physical requirements.
Depending
on the sign of a {\it modified} {\it scattering} {\it length},
it reduces to 
either 
a
{\it collective} {\it rotation} or 
a
{\it condensed} {\it vortex} 
{\it state},
with no alternative. 
The ground state can undergo a kind of quantum phase transition
when
the shape of the interaction potential is smoothly varied.
\\
KEYWORDS: Bose-Einstein condensation, atomic traps, cold atoms,
arbitrary forces, yrast states
\end{abstract}
\pacs{PACS numbers: 03.75.Fi,
32.80.Pj, 67.40.Db, 03.65.Fd
}

Newly developed techniques \cite{STRENTHmanipulating} 
for manipulating the strength of the 
effective interaction between trapped atoms 
\cite{ANDERSON}
open the
possibility to experimentally 
realize a
Bose gas 
with weak interatomic interactions that are either attractive or repulsive.
The limit $\langle$$V$$\rangle$$/$$\hbar$$\omega$$\rightarrow$$\pm$$0$,
where $\langle$$V$$\rangle$ is a typical expectation value
of the interaction and $\hbar$$\omega$ is the quantum 
energy of the confining potential,
is now reached experimentally via the Feshbach resonance
\cite{STRENTHmanipulating}.
Of special interest are the {\it ground} {\it states} of 
a rotating system\cite{MCBD,nature,WGS,REVIEW}
at given angular momenta $L$, 
the so-called {\it yrast states} 
\cite{WGS,REVIEW,Mottelson,BP,PRA1,PRA2}.

So far, the problem was usually studied 
in 
dimension 
$k$$=$$2$ 
for the contact interaction  
$V$$\sim$$a$$^{sc}$$\delta$$($$\vec{r}$$)$,
either attractive ($a$$^{sc}$$<$$0$) or repulsive ($a$$^{sc}$$>$$0$)
where $a$$^{sc}$ is the scattering length.
The 
case 
$a$$^{sc}$$<$$0$ 
was solved analytically 
by Wilkin {\it et al}.\cite{WGS}, 
while the 
case 
$a$$^{sc}$$>$$0$
was studied
numerically 
by Bertsch and Papenbrock
\cite{BP}. 
The conjecture for the ground state wave function\cite{BP}
was confirmed analytically in
\cite{PRA1}, 
and generalized to a universality class of 
predominantly repulsive interactions\cite{PRA1}.
A
universality class of predominantly attractive interactions
has been constructed in\cite{PRA2}.

Since experiments are now feasible \cite{STRENTHmanipulating}
in the 
{\it weak-interaction} {\it limit} 
close to the region where the scattering length changes sign,
it becomes of importance to generalize these results to {\it arbitrary}
interactions.
In this Letter 
the 
problem is solved {\it exactly} in $k$ dimensions
for {\it arbitrary} two-body central forces $V$$($$r$$)$.
The final result is 
the following.
In the 
functional space $\{$$V$$\}$ of all possible interactions $V$$($$r$$)$,
we may restrict our attention to those of physical interest,  $\{V_{phys}\}$,
of which we require that the force $-$$\frac{dV}{dr}$ 
changes sign only once from repulsive at short to attractive at long distance:
\begin{eqnarray}\label{physical}
\frac{dV}{dr} < 0 ,\quad r< R; \qquad \frac{dV}{dr} > 0,  
\quad r\geq R; \qquad R < 1.
\end{eqnarray}
Since the crossover at 
$R$
occurs for atomic reasons, we may assume 
it to be smaller than the trapping size,
$R$$<$$\omega^{-1/2}$ (or $R$$<$$1$ in units $\omega$$=$$1$).
In addition, we assume that the force does not grow at $r$$\rightarrow$$\infty$
and the quantum eigenvalue problem is free of divergencies\cite{WELL}.
It will be shown that
the entire functional space $\{$$V$$_{phys}$$\}$ is divided into two distinct
classes  $\{$$V^-_{phys}$$\}$ and  $\{$$V^+_{phys}$$\}$
of (effectively) attractive or repulsive interactions.
(The meaning of 'effective' in this context will be detailed below.)
Within each class the energies of the yrast states depend in a simple way
on the interaction while their wave functions remain the same.
The 
ground states
in the two classes differ qualitatively.
When the interaction is changed from effectively attractive to repulsive,
the transition between the two
classes can be visualized as a quantum 
phase transition,
with the relative interparticle angular momentum as a discrete order parameter. 
These exact analytical results are exemplified by analysis of Morse
potentials with variable scattering length.

The Hamiltonian of $N$ spinless bosons of mass $m$ in a 
$k$-dimensional symmetric harmonic trap
($k$$=$$2$$,$$3$$,$$.$$.$$.$) 
reads
\begin{eqnarray}\label{HA}
H=
\sum_i^N\left(\frac{\vec{p}_i^2}{2m}+ 
\frac{m\omega^2 \vec{r}_i^2}{2}\right)+
\sum_{i>j}^N V(r_{ij}) \equiv H_0 + V,
\end{eqnarray}
where $H$$_0$ describes noninteracting particles, 
$\vec{r}$$_i$($\vec{p}$$_i$) is the $k$-dimensional position (momentum)
vector of $i$-th boson,
and
$V$
is the two-body interaction  
with $r$$_{ij}$$\equiv$$|$$\vec{r}$$_i$$-$$\vec{r}$$_j$$|$.
Hereafter, $|$$0$$_L$$)$ denotes 
the ground state at fixed $L$ and with maximum value 
of the
conserved component,
$L_{xy}$\cite{noteCOMMENT}.
Other rotationally degenerate wave functions
can be obtained from $|0_L)$ by applying the 
standard angular momentum ladder
operators 
(e.g., $L$$_{-}$$=$$L$$_x$$-$$i$$L$$_y$, in $k$$=$$3$).
In the limit 
$\hbar$$\omega$$\gg$$V$,
the determination of
$|$$0$$_L$$)$ 
requires 
the diagonalization of the interaction 
within the space
\begin{eqnarray}\label{BASIS}
Sz_1^{l_1} z_2^{l_2} . . . z_N^{l_N} |0\rangle,   
\quad
\sum\limits_{i=0}^{N} l_i = L ,
\end{eqnarray}
where  
$z$$_i$$=$$x$$_i$$+$$i$$y$$_i$, $S$ is the symmetrization operator,
and $|0$$\rangle$$\equiv$$e$$^{-\frac{1}{2}\sum\vec{r}_k^2}$,
in the convention $\hbar$$=$$m$$=$$\omega$$=$$1$.
The states involving 
$z^*_i$$=$$x$$_i$$-$$i$$y$$_i$
are separated from (\ref{BASIS}) 
by an energy $\delta$$E$$=$$n$$\hbar$$\omega$
and  their admixtures  can be neglected  
for 
$V$$\ll$$\hbar$$\omega$.
The 
dimensionality of the basis 
(\ref{BASIS}) 
grows exponentially with $L$\cite{Mottelson}.

Within the 
subspace (\ref{BASIS}) the Hamiltonian 
(\ref{HA})
is
\begin{eqnarray}\label{ham}
H= 
L+(Nk)/2 + W, 
\end{eqnarray}
where
the first (constant) terms come from $H$$_0$
and 
$W$ is the interaction $V$, 
projected \cite{PRA1} onto the
subspace  (\ref{BASIS}).
Using the integral transform 
\begin{eqnarray}\label{MELLIN}
w(l) \equiv \int_0^{\infty} 
V(\sqrt{2t})
e^{-t} 
\frac{t^{l+{k}/{2}-1}}{\Gamma(l+k/2)}  dt,
\end{eqnarray}
we can write 
$W$
in the form
\begin{eqnarray}\label{Hrepresentation}
W = S\sum_{i>j} w(\hat{l}_{ij})S ,
\quad
\hat{l}_{ij} =(a^+_i - a^+_j)(a_i - a_j)/2
\end{eqnarray}
with $\hat{l}_{ij}$ the relative angular momentum 
between atoms  $i$ and $j$ written in terms of the ladder operators
$a^+_i$$=$$\frac{z_i}{2}$$-$$\frac{\partial}{\partial z^*_i}$
and $a_i$$=$$\frac{z^*_i}{2}$$+$$\frac{\partial}{\partial z_i}$.
The
total {\it internal angular momentum}
$J$$=$$\sum_{i>j}$$\hat{l}$$_{ij}$
is an exactly conserved quantity with eigenvalues 
$J$$=$$\frac{Nj}{2}$, $j$$=$$0$$,$$2$$,$$3$$,$$.$$.$$,$$L$\cite{PRA1},
commuting with 
the total angular momentum 
$L$$\equiv$$\sum_i$$a^+_i$$a_i$.

We 
split
\cite{PRA1} 
the 
projected interaction
$W$ 
into 
$W$
$=$
$W$$_0$
$+$
$W$
$_S$
such that the first term is simple enough to find 
its lowest eigenvalue $E$$_0$ and its associated eigenstate $|$$0$$)$,
\begin{eqnarray}\label{SUSY0}
W_0 |0) = E_0 |0) .
\end{eqnarray}
The state  $|0)$ will also be the ground state of the total interaction  
$W$$=$$W$$_0$$+$$W$$_S$
if (i) 
$W_S$
is  {\it non-negative definite}, 
\begin{eqnarray}\label{SUSY2}
W_S \equiv S \sum_{i>j}v_S(ij) S
\geq 0 ,
\end{eqnarray}
and (ii) $|0)$ is annihilated by   
$W_S$\cite{PRA1}, 
\begin{eqnarray}\label{annihilation}
W_S|0)=0 .
\end{eqnarray}
In general, the operator 
$W_0$
can be written as
\begin{eqnarray}\label{ansatz}
W_0 =  S\sum_{i>j} v_0(\hat{l}_{ij})S, 
\quad v_0(l)=  
\sum\limits_{m=0}c_m l^m  ,
\end{eqnarray}
where  $c$$_m$ are hitherto unknown coefficients that need to be fixed such 
that conditions (\ref{SUSY0}), (\ref{annihilation}), and (\ref{SUSY2})
are satisfied.
The condition (\ref{SUSY2}) can be simplified in the following way.
The operator 
$v_S(ij)$ can be diagonalized
in the 
unsymmetrized monomial
states in Eq.(\ref{BASIS}) 
$|m\rangle\equiv z_1^{l_1}
z_2^{l_2} . . . z_N^{l_N}|0\rangle$  via the substitution
$z_i$$\rightarrow$$\frac{z_i-z_j}{\sqrt{2}}$, $z_j$$\rightarrow$$\frac{z_i+z_j}{\sqrt{2}}$. 
Its
eigenvalues are 
$w$$($$l$$)$$-$$v$$_0$$($$l$$)$, 
$l$$=$$0$$,$$1$$,$$2$$,$$.$$.$$.$$,$$L$\cite{noteHALDANE}. 
Odd-$l$
eigenvectors are annihilated by $S$.
Therefore, our strategy is to 
choose
the $c$$_m$ in such a way that 
some of the even-$l$ eigenvalues
$\lambda$$_n$$=$$w$$($$2$$n$$)$$-$$v$$_0$$($$2$$n$$)$
are zero and all others positive.
This will
guarantee
that 
$W$$_S$ is a non-negative definite operator\cite{PRA1},
even though
we cannot evaluate its eigenvalues directly.
The candidates for the state $|$$0$$)$ satisfying (\ref{SUSY0}) and (\ref{annihilation})  
can be sought among 
linear combinations of 
states
$S$$|$$m$$\rangle$'s which obey $v$$_S$$|$$m$$\rangle$$=$$0$.

The
following solution for the $c$$_m$ satisfies 
(\ref{SUSY0}), (\ref{SUSY2}), 
(\ref{annihilation}):
\begin{equation}\label{solution}
c_0 = w(0), \qquad c_1 = -\Delta_2 \theta(\Delta_2)/2 ,  \qquad c_{m\geq 2} =0 ,
\end{equation}
where  
$\theta(x)=\{0(x\leq 0), 1(x>0)\}$ 
is the 
step function and $\Delta$$_{2n}$$\equiv$$w$$($$0$$)$$-$$w$$($$2$$n$$)$.
The eigenvalues 
$\lambda^{\pm}_n$ 
(for $\Delta$$_2$$\stackrel{\textstyle{>}}{\textstyle{<}}$$0$)
of the corresponding operator $v$$_S$$($$i$$j$$)$ in (\ref{SUSY2}) read
\begin{equation}\label{VS}
\lambda_n
= n \theta(\Delta_2)\Delta_2 - \Delta_{2n} \geq 0 .
\end{equation}
As will be shown below, 
the inequalities are satisfied
for all interactions $V$$_{phys}$
defined in 
(\ref{physical}).  
The ground state and its energy as a function of $L$ ($L$$\leq$${N}$)
are
\begin{eqnarray}\label{RESULT}
|0_{L}) = e^{-\frac{1}{2}\sum\vec{r}_i^2} 
S
\prod_{j=1}^{L}\sum\limits_{i=1}^{N}(z_i-{z}_j\theta(\Delta_2)) ,
\\
E_0(L)= L+\frac{Nk}{2}+ \frac{N(N-1)w_0}{2} - 
\frac{\epsilon_L NL\Delta_2\theta(\Delta_2)}{4}
.\nonumber
\end{eqnarray}
Here 
$\epsilon$$_L$$=$$\theta$$($$L$$-$$1$$)$. 
The ground state depends on 
the interaction $V$$($$r$$)$
only via the {\it sign} of the 
control parameter
\begin{eqnarray}\label{delta2}
\Delta_2=
\frac{2^{1-{k}/{2}}}{\Gamma({k}/{2})}
\int_0^{\infty} V(r) 
r^{k-1} e^{-\frac{r^2}{2}} \left(1-\frac{r^4}{k^2+2k}\right) dr.
\end{eqnarray}
Among the potentials  
$\{$$V$$_{phys}$$\}$, 
the equation
\begin{eqnarray}\label{separatrix}
\Delta_2=0 
\end{eqnarray}
selects
the submanifold 
$\{$$V^0_{phys}$$\}$ 
which we call {\it separatrix}.
The  {\it separatrix} divides 
the interactions 
$V$$_{phys}$$($$r$$)$ 
into the two classes  
$\{$$V^-_{phys}$$\}$ and  $\{$$V^+_{phys}$$\}$,
differing by the form of the ground state:
If $\Delta$$_2$$\leq$$0$,
the ground state
(\ref{RESULT})
becomes
\begin{eqnarray}\label{RESULT2attractive}
|0^{-}_{L}) 
=
e^{-\sum\frac{\vec{r}_i^2}{2}}  
Z^L,
\qquad
\hat{l}_{ij} |0^-_{L}) = 0,
\end{eqnarray}
where $Z$$\equiv$$\sum_i^N$$\frac{z_i}{N}$ 
\cite{noteYRASTlimit}
while 
for $\Delta$$_2$$\geq$$0$ and 
$L$$>$$1$
(\ref{RESULT}) gives
\begin{eqnarray}\label{RESULT2}
|0^{+}_{L}) 
=
e^{-\sum\frac{\vec{r}_i^2}{2}} 
\hat{S} \bar{z}_1 \bar{z}_2...  \bar{z}_L , 
\qquad J |0^{+}_{L}) = \frac{NL}{2} |0^{+}_{L}).
\end{eqnarray}
with 
$\bar{z}$$_j$$=$$Z$$-$$z$$_j$. 
The internal angular momentum
$J$
can be 
viewed as
a ``discrete order parameter'',
zero in the domain of 
potentials 
$V$$($$r$$)$
with $\Delta$$_2$$\leq$$0$ and positive in the
domain $\Delta$$_2$$\geq$$0$.
For $\Delta$$_2$$\ne$$0$,  (\ref{RESULT2attractive}) and (\ref{RESULT2})
are non-degenerate ground states;
within the {\it separatrix},  
they are degenerate\cite{noteGENyrast}.

For $k$$=$$2$, we have 
$\vec{r}^2_i$$=$$|$$z$$|_i^2$ in Eqs.(\ref{RESULT2attractive},\ref{RESULT2})
and
the $|$$0^{\mp}_L$$)$ 
accord with 
the results
obtained for the attractive\cite{WGS} and repulsive\cite{BP},\cite{PRA1}
contact interactions, respectively. 
We arrive at an interesting conclusion:
exact solution for the problem (\ref{HA}) for 
{\it arbitrary} {\it interaction} is given by simple generalization
of the results for the universality classes of 
{\it predominantly} {\it attractive} \cite{PRA2} 
and {\it repulsive} \cite{PRA1} interactions.

By $\nu^{\pm}_n$ we denote the normalized occupancies $\nu$$_n$ ($\sum_n$$\nu$$_n$$=$$1$)
of the single-particle 
oscillator states $z$$^n$
in the ground 
states $|$$0$$^{\pm}$$)$.
 The two behave qualitatively differently.
For $L$$=$$0$ we have $\nu^+_n$$=$$\nu^-_n$$=$$\delta$$_{n,0}$, that is the trivial
(noninteracting) condensate.
For $1$$<$$L$$<N$ we have
\begin{displaymath}
\nu^+_n = 
\frac{L![q(\alpha-1)! {_2}F_0(\alpha,\gamma;s_1) + t{\alpha}!{_2}F_0(\alpha+1,\gamma;s_1) ]}
{s_1^3N^{2n+4}n!(-\gamma)!(N-L)!{_2}F_0(-L,N-L+1;s_0) }. 
\end{displaymath}
Here ${_2}$$F$$_0$ is the hypergeometric sum\cite{ABR},
$q$$=$$\alpha$$+$$N$$\beta$$^2$$s$$_1$
and 
$t$$=$$s$$_1$$($$2$$\beta$$+$$1$$)$$-$$1$
with
$s$$_i$$=$$($$i$$+$$N$$)$$/$$N$$^2$,
$\alpha$$=$$N$$+$$\gamma$,
$\gamma$$=$$n$$-$$L$,
and 
$\beta$$=$$N$$-$$L$$-$$N$$n$.
As $L$ grows,
the second excited $z$$^2$ and the first excited $z$$^1$
states
become populated (see Fig.1.).
For $L$$\ne$$0$$,$$N$, the $\nu^+_0$, $\nu^+_1$ and $\nu^+_2$ remain macroscopic
[O(1)] 
in the limit 
$N$$\gg$$1$.
For $L$$\rightarrow$$N$ 
the bosons condense in the state $z$$^1$,
forming a vortex. 
We 
call
$|$$0$$^+_L$$)$ 
{\it condensed vortex state}\cite{PRA1}.
In contrast, the state $|$$0^-_L$$)$ 
gives 
a binomial distribution 
$\nu^-_n$ $=$ $\frac{L!(N-1)^{L-n}N^n}{(L-n)!n!(2N-1)^L}$
peaked at 
$n$$=$$\frac{L}{2}$\cite{WGS}.
We call $|$$0^-_L$$)$ a {\it collective} {\it rotation} state:
its angular momentum comes from the collective factor $Z$$^L$.
An interesting 
way to view 
such states as 
rotated
noninteracting condensates was 
presented
in \cite{PP}.

The control parameter  $\Delta$$_2$ measures the balance between repulsion
and attraction in $V$$($$r$$)$. 
For zero-range interactions,  
$\frac{\Gamma(k/2)}{2^{1-k/2}}$$\Delta_2$ 
coincides 
with the scattering length in Born approximation,
$a^{sc}_k$$=$$\int_0^{\infty}$$V$$($$r$$)$$r$$^{k-1}$$d$$r$.
For finite-range interactions it differs from $a^{sc}_k$; 
and it
can be considered as a {\it modified} {\it scattering} {\it length}
which separates the two universality classes of
{\it effectively} {\it attractive} and {\it effectively} {\it repulsive} 
interactions.
%
%

For $\Delta$$_2$$\leq$${0}$, the interaction contribution
to the ground states energy 
(\ref{RESULT}) is negative
and $L$-independent. 
It grows with $\Delta$$_2$.
At the critical point $\Delta$$_2$$=$$0$, the derivative 
$\partial$$E$$_0$$($$L$$)$$/$$\partial$$\Delta$$_2$
has a jump $\frac{-NL\epsilon_L}{4}$ and $E$$_0$$($$L$$)$ 
becomes $L$-dependent for $\Delta$$_2$$\geq$$0$.  
In Fig.2, we show the
results for the
Morse potential 
$V$$_M$$($$r$$)$$=$ $e$$^{\frac{2(R-r)}{a}}$ $-$$2$$e^{\frac{R-r}{a}}$.
We display $V$$_M$
for two sets of parameters $a$ and $R$, 
and the resulting
$l$-even 
part, $w$$_e$,
of
the transform (\ref{MELLIN})
\begin{eqnarray}\label{Wpotentials}
w(l) 
= 
\Gamma(l+\frac{k+1}{2})\frac{2^{l+\frac{k}{2}}}{\pi^{1/2}}
\left[e^{\frac{2R}{a}}d_{-2l-k}^{(2/a)}-2e^{\frac{R}{a}}d_{-2l-k}^{(1/a)}
\right] ,
\end{eqnarray} 
where
$d$$_l$$^{(s)}$$=$$e$$^{s^2/4}$$D$$_{l}$$($$s$$)$ 
with $D_l$$($$s$$)$ 
the parabolic cylinder function\cite{ABR}.
It is seen that 
$\lambda$$_n$$\equiv$$w$$($$2$$n$$)$$-$$v$$_0$$($$2$$n$$)$$\geq$$0$ 
in both cases.
When
attraction prevails, $\Delta$$_2$$<$$0$, 
$w$$_e$ has a minimum at 
$l$$=$$0$
and
the system resides in its lowest-energy state 
$|$$0^-_L)$
for a minimum possible value of 
$J$$=$$\sum$$l$$_{ij}$$($$=$$0$$)$.
When repulsion starts to prevail,
the quantity 
$\Delta$$_2$
becomes positive, the minimum of $w$$_e$ is shifted to 
$l$$>$$0$, 
and the system finds 
a lower energy for 
a 
non-zero value of 
$J$
in the ground state  
$|$$0^+_L)$.
The behavior of $w$$_e$$($$l$$)$ at the critical point $\Delta$$_2$$=$$0$
resembles that of a thermodynamic potential in a second-order 
phase transition 
\cite{LANDAU}.
In contrast to thermodynamical systems, this ``phase transition'' is insensitive to 
$k$ and it happens for finite $N$. 
Panel c) shows the bifurcation point in $E$$_0$$($$L$$)$.

Fig.2 d) shows the phase diagram for the Morse 
potentials
in the parametric space $($$a$$,$$R$$)$.
The separatrix (\ref{separatrix}) between 
$\{$$V$$^-$$\}$ and $\{$$V$$^+$$\}$
reduces to a curve
defined
by
the following relation between $a$ and $R$
\begin{eqnarray}\label{SHORTseparatrix}
R(a)=a\ln(\phi^{(1/a)}_{-k-1}/\phi^{(2/a)}_{-k-1}), \quad
\phi^{(s)}_k{\equiv}{d_{k}^{(s)}}-kd_{k-2}^{(s)}.
\end{eqnarray}
%
%
%
Using Eqs.(\ref{Wpotentials},\ref{SHORTseparatrix}),
$W$$_S$ can be shown to be
non-negative definite (\ref{VS}) throughout:
For values of 
$R$$\leq$$R$$($$a$$)$,
we obtain
$\Delta$$_2$$\leq$${0}$
and 
$\lambda^-_n$$\geq$$0$;
for 
$R$$\geq$$R$$($$a$$)$, 
we have
$\Delta$$_2$$\geq$${0}$ and 
$\lambda^+_n$$\geq$$0$.
Thus {\it all} Morse potentials, whatever their values of $a$ and $R$
are covered by (\ref{RESULT}) and
fall into two distinct classes depending on whether $\Delta$$_2$ is positive or
negative.
[Note that for $R$$>$$1$, the $V$$_M$ 
lies outside the class (\ref{physical})!] 
We obtained similar results for other two-parametric potential families.

Is this situation generic? 
The 
full
set
of physical interactions (\ref{physical})
forms a dense functional manifold. In general, it cannot be 
described by a finite number of parameters. 
Let 
$\{$$V$$\}$
be the complete manifold of all 
potentials
[not necessarily the physical potentials (\ref{physical})]
as
indicated
by the big 
disc
in Fig.3. 
Within this extended manifold,
we can still define the subclasses 
that have
$\lambda^-_n$$\geq$${0}$ and $\lambda^+_n$$\geq$${0}$
with the ground states 
(\ref{RESULT2attractive}) and (\ref{RESULT2}), respectively.
Their boundaries $\Lambda$$^-$ and $\Lambda$$^+$
can in general be distinct,
leaving room 
marked by $''?''$ 
when the ground state 
is not (\ref{RESULT}).
We will 
show that there is no such gap 
for all the physical interactions (\ref{physical}),
confirming that the result (\ref{RESULT}) covers the whole 
{\it universality class} 
$\{$$V$$_{phys}$$\}$
and that
the separatrix $\Delta$$_2$$=$$0$ (\ref{separatrix}) 
divides 
the interactions 
$\{$$V$$_{phys}$$\}$ 
(small 
disc
in Fig.3.) into
the two classes, with the ground states 
(\ref{RESULT2attractive}) and (\ref{RESULT2}),
respectively, with no other alternative.
The proof requires to show that for any potential 
in $\{$$V$$_{phys}$$\}$ 
$\lambda^-_n$$\geq$$0$ for $\Delta$$_2$$\leq$$0$  
and $\lambda^+_n$$\geq$$0$ for $\Delta$$_2$$\geq$$0$,
Eq.(\ref{VS}).

%
%

The quantity $\Delta$$_{2n}$$=$$\int_0^{\infty}$$\phi_{2n}$$F$$d$$t$ 
is the work 
done 
by the modified force 
$F$$\phi$$_{2n}$$\equiv$$\frac{-dV(\sqrt{2t})}{dt}$$\phi$$_{2n}$
to separate a pair of particles.
Here 
$\phi$$_{2n}$$\equiv$$\sum_{m=1}^{2n}$$\frac{e^{-t}t^{k/2+m-1}}{\Gamma(k/2+m)}$
is positive definite.
By (\ref{physical}) the work is the sum of the positive and negative areas 
\begin{eqnarray}\label{DELTA}
\Delta_{2n} =  \int_0^{\tau} h_{n}\phi_2 F dt +  
\int_{\tau}^{\infty} h_{n} \phi_2 F dt ,
\qquad \tau \equiv \frac{R^2}{2} .
\end{eqnarray}
where $h_n$$\equiv$$\frac{\phi_{2n}}{\phi_2}$ 
are positive-valued functions.
If $\Delta$$_2$$\leq$${0}$, 
the second (negative) term prevails for $n$$=$$1$.
It will then prevail for any $n$$>$$1$, because 
the functions
$h$$_{n>1}$ 
increase monotonically. 
That is, if 
$\Delta$$_2$$\leq$$0$, then all 
$\Delta$$_{2n}$$\leq$$0$ and $\lambda^-_n$$\geq$$0$.
If $\Delta$$_2$$\geq$${0}$, 
the first (positive) term 
in (\ref{DELTA}) prevails for $n$$=$$1$. 
We introduce 
the monotonically decreasing functions $\tilde{h}$$\equiv$$n$$-$$h$$_n$.
They
are positive for $t$$<$$t$$_n$ and negative for $t$$>$$t$$_n$ with all $t$$_n$$>$$1$. 
By (\ref{physical})
we have
$\tau$$<$$t$$_n$ and we write
\begin{displaymath}
\lambda^+_n
= 
\int_0^{\tau} \tilde{h}_n \phi_2 F dt + 
\int_{\tau}^{t_n} \tilde{h}_n \phi_2 F dt + 
\int_{t_n}^{\infty}  \tilde{h}_n \phi_2 F dt.
\end{displaymath}
The second term is the only negative contribution. 
The corresponding area is however smaller than that of the first term, 
because $\tilde{h}$$_n$ decreases monotonically and $\tau$$<$$t$$_n$. 
Thus $\lambda^+_n$$\geq$$0$ 
for  $\Delta$$_2$$\geq$${0}$.
To sum up, the inequalities
(\ref{VS})
hold throughout,
and the solution (\ref{RESULT}) covers the universality class $\{$$V$$_{phys}$$\}$.

By similar arguments, one can append 
the class 
$\{$$V$$_{phys}$$($$r$$)$$\}$ (\ref{physical})
by potentials with 
constant sign of $F$ and 
$\frac{d^2V(\sqrt{2t})}{dt^2}$, 
like
$\delta$$($$\vec{r}$$)$,
$\frac{1}{r}$, 
$l$$o$$g$$($$r$$)$, 
$e$$^{-r/a}$, $\frac{e^{-r/a}}{r}$,
${e^{-r^2/a^2}}$ {\it etc}.

In summary, we have solved the problem of
rotating ground states of weakly interacting trapped Bose atoms.
The resulting 
phase diagram gives a 
complete classification of these ground states.
The conclusion is that the 
ground state is either a collective rotation or a condensed vortex state;
in both cases analytical expressions are known for the ground-state wave function
and its energy and these are valid whatever the form of interatomic interaction,
as long as it has reasonable properties.
These results can be of interest in view of recent experiments with trapped atoms
and can be applied to other systems.

The work was supported by CEA (France) and FAPESP (Brazil).

\newpage

{\large Figure Captions}

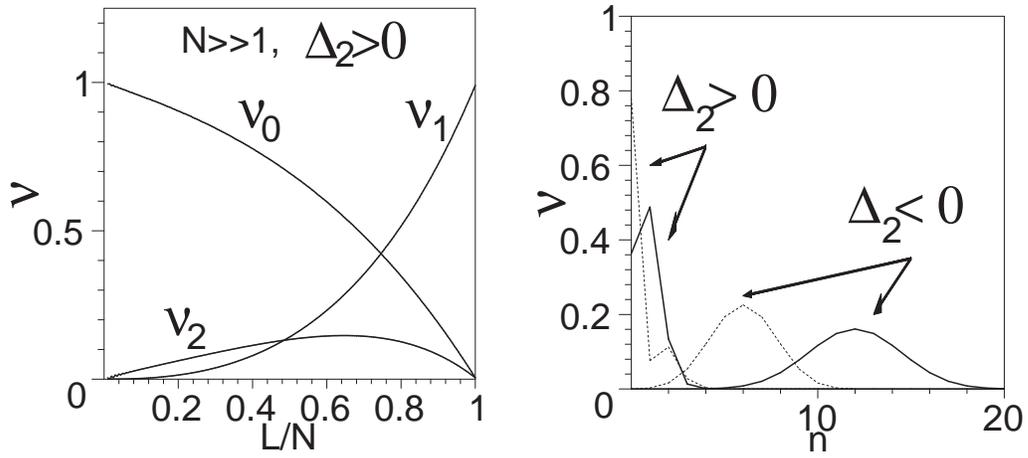
FIG. 1.
Left: Asymptotic values of $\nu^+_n$ as functions of $L$$/$$N$ for 
$n$$=$$0$$,$$1$$,$$2$ at 
$N$$\gg$$1$.
Right: $\nu$$_n$ 
as functions of $n$
for $L$$=$$1$$2$ (dash) and
for  $L$$=$$2$$4$ (solid) for $\Delta$$_2$$>$$0$ and $\Delta$$_2$$<$$0$ ($N$$=$$3$$0$).

\vspace{5mm}

FIG. 2.
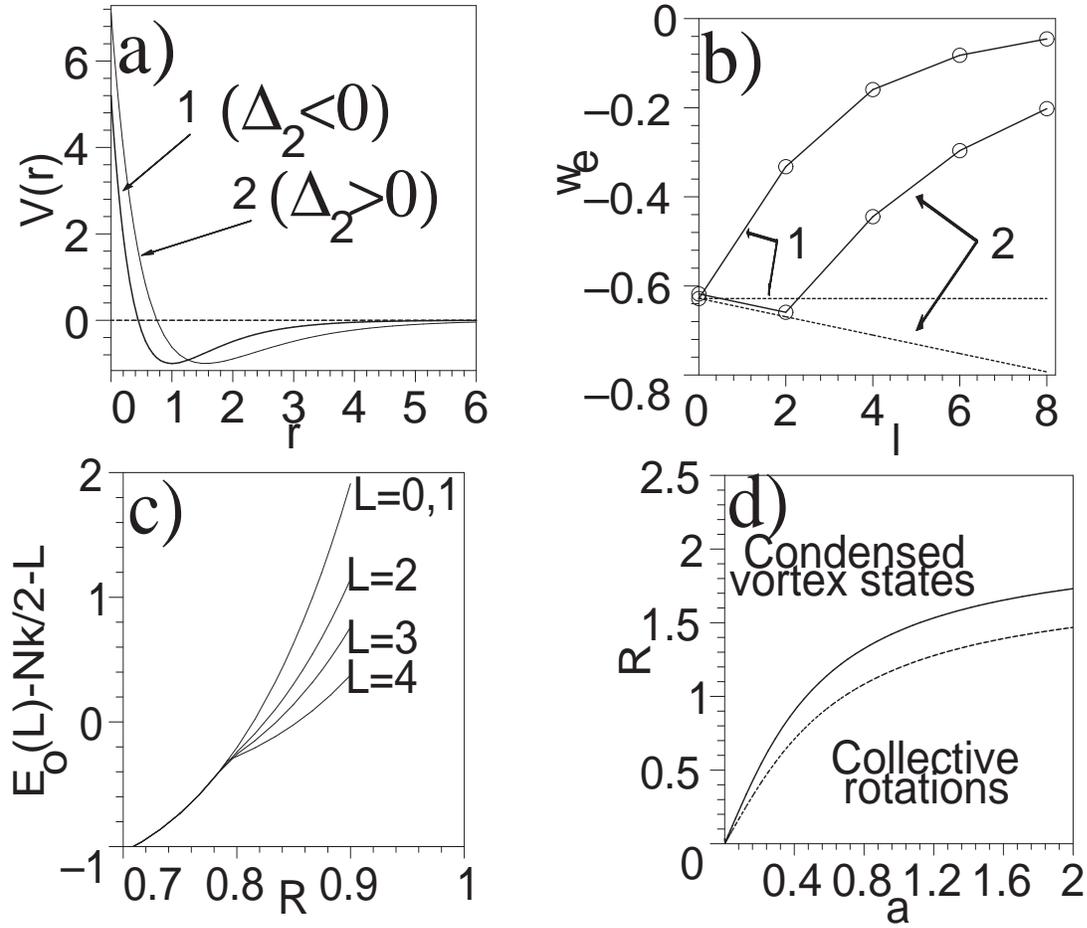
Morse
potential $V$$_M$$($$r$$)$ (a) and 
$w$$_e$$($$l$$)$
from (\ref{Wpotentials})  (b)
for the two sets of parameters: ``1''($a$$=$$0$$.$$8$,$R$$=$$1$$.$$0$,$\Delta$$_2$$>$$0$)
and ``2''($a$$=$$1$$.$$1$$5$,$R$$=$$1$$.$$5$$5$,$\Delta$$_2$$>$$0$), $k$$=$$3$.
Symbols connected by solid lines show $w$$_e$.
The $v$$_0$ (\ref{ansatz},\ref{solution}) are shown by dashed lines.
(c)
The interaction contribution to the ground state energy 
(\ref{RESULT2}), 
as function of $R$ (for $a$$=$$1$$/$$3$, $k$$=$$3$ and $N$$=$$4$). 
(d) 
Phase diagram for family of the Morse 
potentials
on the $a$$-$$R$-plane.
The separatrices (\ref{SHORTseparatrix}) 
for $k$$=$$3$ (solid) and $k$$=$$2$ (dash).

\vspace{5mm}

FIG. 3.
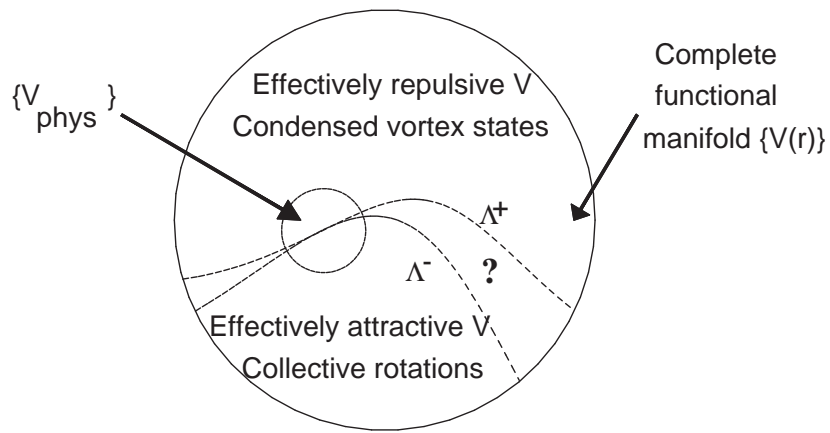
Schematic {\it phase diagram} in functional space $\{$$V$$($$r$$)$$\}$.
Within $\{$$V$$_{phys}$$\}$, the boundaries $\Lambda$$^-$ and $\Lambda$$^+$
merge with the {\it separatrix} $\Delta$$_2$$=$$0$ (\ref{separatrix}).


\begin{thebibliography}{300}
\bibitem{WELL} In practice, this implies that $V$$($$r$$)$ does not grow
as (or faster than) $r$$^{-k}$ at $r$$\rightarrow$$0$.
\bibitem{noteCOMMENT}
That is, 
$L_{xy}$$|$$0$$_L$$)$$=$$L$$|$$0$$_L$$)$  
and  $\vec{L}$$^2$$|$$0$$_L$$)$$=$$L$$($$L$$+$$k$$-$$2$$)$$|$$0$$_L$$)$ 
with 
$\vec{L}$$^2$$\equiv$$\sum_{\alpha>\beta}^k$
$($$r$$^{\alpha}$$p$$^{\beta}$$-$$r$$^{\beta}$$p$$^{\alpha}$$)$$^2$
where $\alpha$$,$$\beta$$=$$1$$,$$.$$.$$,$$k$ denote the spatial components.
\bibitem{noteYRASTlimit}
The limitation $L$$\leq$$N$
does not apply in the $\Delta$$_2$$<$$0$ case.
\end{thebibliography}
\end{document}